\documentclass[aps,twocolumn,showpacs]{revtex4}

\usepackage{graphicx}
\usepackage{amsmath}

\def\fig#1{Fig.\,\ref{#1}}
\def\tab#1{Tab.\,\ref{#1}}
\def\eq#1{Eq.~(\ref{#1})}

\def\beq#1{\begin{equation}\label{#1}}
\def\eeq{\end{equation}}
\def\width{L}
\def\aphi#1{a_{\varphi_#1}}
\def\bphi#1{a'_{\varphi_#1}}
\def\aphi#1{A_{#1}}
\def\bphi#1{A'_{#1}}
\def\vec#1{\mathbf{#1}}
\def\i{\mathrm{i}}
\def\f{\mathrm{f}}

\begin{document}
\title{Rescattering for extended atomic systems}
\author{Ulf Saalmann and Jan M. Rost}
\affiliation{Max Planck Institute for the 
  Physics of Complex Systems\\
  N\"othnitzer Stra{\ss}e 38, 01187 Dresden, Germany}
\date{\today}

\begin{abstract}\noindent 
  Laser-driven rescattering of electrons
  is the basis of many strong-field phenomena in atoms and
  molecules.  
  Here, we will show how this mechanism operates in extended
  atomic systems, giving rise to effective energy absorption.  
  Rescattering from extended systems can also lead to energy
  loss, which in its extreme form results in non-linear
  photo-association. 
  Intense-laser interaction with atomic 
  clusters is discussed as an example.
  We explain fast electron emission, seen in experimental
  and numerically obtained spectra, by rescattering of
  electrons at the highly charged cluster.
\end{abstract}


\maketitle

\noindent 
Laser-driven rescattering of electrons \cite{co93} is at the
heart of strong-field atomic physics.  
The basic principle is very simple: 
a bound electron is released from an atom (or a negative ion
\cite{frfl+05}) with the help of the strong electric field of a
laser, by which it is subsequently driven back to the ion.  
On return to the ion the
electron may recombine (emitting harmonic radiation \cite{leba+94}
with the access to attosecond laser pulses \cite{sciv+06}), ionize
other electrons (inducing multiple ionization \cite{bedo+05}) or may
be backscattered (gaining high kinetic energy termed above-threshold
ionization \cite{mipa+06}).   Due to the strong dependence of the
tunnel probability on the field the release time is restricted to
phases of the laser period with maximal  electric field.
In the following we concentrate on linearly polarized light 
which exhibits the most pronounced rescattering effects.

Most direct evidence for the rescattering mechanism comes from
measuring kinetic energies of the released electrons
\cite{agfa+79,cobu+89,pani+94,wash+96}.
The  momentum $p$ an electron can acquire in an
oscillating field $f(t)=F\cos(\omega t)$ of strength $F$ and
frequency $\omega$ depends on the phase $\varphi_0=\omega t_0$ at the
time $t_0$ of its release. At this time the vector potential of the 
field is $\aphi{0}:=A\sin\varphi_0\equiv(F/\omega)\sin\varphi_0$ and this 
is exactly the momentum acquired, $p = \aphi{0}$.
%
(We use atomic units.)  
Obviously, the maximum 
$p_\mathrm{max}=A\equiv F/\omega$ 
occurs at $\varphi=\pi/2$ and
results in a kinetic energy $E_\mathrm{max}=2E_\mathrm{pond}$
with the ponderomotive energy $E_\mathrm{pond}:=F^2/4\omega^2$.
However, electrons with such high energies are rare
\cite{cobu+89} since they must be released from the atom when
the electric field vanishes, which is very unlikely (see above).  
Electrons with energies even beyond $2E_\mathrm{pond}$ are
indicative for rescattering of electrons  \emph{previously} released.

For atoms the rescattering process can be understood in classical
terms by an electron elastically 
\footnote{The collision with the
potential itself is elastic, i.\,e., does not change the energy
of the electron.  It can, however, absorb energy from the
time-dependent field.}
scattered at a zero-range potential in the
presence of an electromagnetic field \cite{pabe+94}; neglecting the
Coulomb tail turned out to be of minor importance.  In this case, and
more generally, if the scattering time is much shorter than the
laser period, the absorbed energy $\Delta E$ is given by \cite{krwa73}
\begin{equation}
  \label{eq:abkw}
\Delta E=2\Delta\vec{p}\cdot\vec{A}(t')  
\end{equation}
with $\Delta\vec{p}$ the momentum change at the time $t'$ of scattering.
The absorption is particularly effective for a large momentum
change $\Delta\vec{p}$, which is achieved through
\emph{backscattering}, 
and for scattering events taking place at a maximal vector
potential (\emph{minimal\/} electric field).

For an extended scattering potential we will see that optimum 
energy absorption upon rescattering requires completely different 
conditions, namely
\emph{forward-scattering\/} at \emph{maximal\/} electric field.

\begin{figure}[b]
  \centering
  \includegraphics[width=0.7\columnwidth]{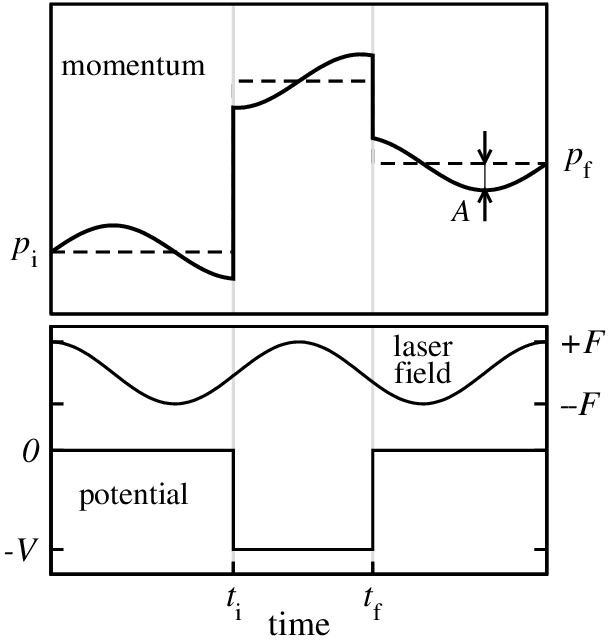} 
  \caption{Schematic picture of the scattering process.
    The upper part shows the electron's momentum,
    the lower part the potential and the value of
    the electric field 
    (assumed to be homogeneous in space)
    as a function of time, respectively.}
  \label{fig:model}
\end{figure}%
This becomes obvious by analyzing the simplest situation of scattering 
from an extended potential. 
It can be  realized with an electron passing under the influence
of an oscillating electric field centrally a spherical well.
The one-dimensional problem is sketched in \fig{fig:model}.  
Assuming a constant potential $-V<0$ inside  the extended
scatterer  naturally generalizes the zero-range potential used
in the context of rescattering at atoms \cite{pabe+94}.  
Note, that similar to the discussion before \eq{eq:abkw} for the
atoms, the time of crossing the potential \emph{boundaries} may
assumed to be short compared to the laser period.  
Therefore, the boundaries are idealized here as steps in the
potential.  
This allows for the corresponding electron dynamics to be solved
analytically.
The electron momentum 
\beq{momentum}
P_{\alpha}=p_{\alpha}+(F/\omega)\cos(\omega t)
\eeq%
contains a constant
drift $p_{\alpha}$ (shown by dashed lines in \fig{fig:model}) 
and an oscillating part due to the laser field (shown by thick
solid lines in \fig{fig:model}).  
In order to conserve energy across the boundaries of the
potential reached at times $t_\i$ and $t_\f$,  respectively,
the electron momentum must jump at those times.

The motion is  determined by the laser (with field strength
$F$ and frequency $\omega$), the potential (with width
$\width$ and depth $V$), and the initial drift momentum $p_\i$, 
cf.\ Fig.\,\ref{fig:model}.
A given entrance time $t_\i$ (or phase $\varphi_\i:=\omega t_\i$)
fixes the exit time $t_\f$  and phase $\varphi_\f:=\omega t_\f$.
Then the final momentum, cf.\ \eq{momentum}, reads
\begin{equation}
  \label{eq:vdve}
  P_\f=
  \left[\left(\aphi{\f}-\aphi{\i}
      +\sqrt{P_\i{}^2+p^2}\right)^2-p^2\right]^{1/2},
\end{equation}%
whereby the phases $\varphi_\i$ and $\varphi_\f$ are connected by
\begin{equation}
  \label{eq:p1p2}
  \width\omega=\left(-\aphi{\i}
    +\sqrt{P_\i{}^2+p^2}\right)(\varphi_\f-\varphi_\i)
  -\left(\bphi{\f}-\bphi{\i}\right).
\end{equation}%
In addition to $\aphi{\alpha}=A\sin\varphi_\alpha$, 
we have defined $\bphi{\alpha}:=A\cos\varphi_\alpha$. 
The ``transit''  momentum
$p:=\sqrt{2V}$ characterizes the depth of the potential. 
We assume $p_\i\ge A$ to guarantee monotonic
motion of the electron and avoid multiple passing of the
potential border.

The final momentum $p_\f$ cannot be expressed explicitely in terms
of $\varphi_\i$ since Eq.\,(\ref{eq:p1p2}) is essentially non-algebraic.
However, it is instructive to optimize $p_\f$ with respect to
$\varphi_\i$ and $\varphi_\f$.  Then, Eq.\,(\ref{eq:p1p2}) reveals the optimum 
width of the potential for extremal energy absorption under  given 
laser light.
One obtains extrema for $\varphi_{\i,\f}=\pm\pi/2\mod2\pi$ and
the maximum $p_\f$ for $\varphi_\i=-\pi/2$ and $\varphi_\f=+\pi/2$ with the electron
passing the center of the potential  at times between $-\pi/\omega$ and
$+\pi/\omega$.  Electron momentum and potential width read
\begin{eqnarray}
  p_{\f,\mathrm{max}} &=&
  -A+\left[\left(2A
      +\sqrt{\left(p_\i-A\right)^2+p^2}\right)^2-p^2\right]^{1/2}
  \label{eq:vmax}
  \\
  \width_\mathrm{max} &=& \frac{\pi}{\omega}
  \left(A+\sqrt{\left(p_\i-A\right)^2+p^2}\right).
  \label{eq:rmax}
\end{eqnarray}
In  the limit of a deep potential,
i.\,e., $p\,{\gg}\,p_\i$  and $p\,{\gg}\,A$,
both expressions simplify  to
\begin{eqnarray}
  p_{\f,\mathrm{max}} &\approx&
  2\sqrt{A\,p}
  \label{eq:vmax1}
  \\
  \width_\mathrm{max} &\approx& p\frac{\pi}{\omega}.
  \label{eq:rmax1}
\end{eqnarray}
Obviously the optimal potential width $\width$ in
Eq.\,(\ref{eq:rmax1}) corresponds to the distance an electron travels
with momentum $p$ during half a cycle $\pi/\omega$ of the laser pulse
\cite{taan+04}.

At optimal width $L_{\mathrm{max}}$ the maximum electron momentum achievable 
depends on both, the laser field amplitude $A$ and the depth of the 
potential (through $p$ in 
Eq.\,(\ref{eq:vmax1})). 
This is different from above-threshold ionization of single atoms
where the maximal electron momentum is  determined by the
laser field only, namely  $p_\mathrm{max}=\sqrt{5}\,A$
(or $E_\mathrm{kin,max}=10E_\mathrm{pond}$) \cite{mipa+06}. 
It should be mentioned that  electrons with such
high energies are typically a few orders of magnitude less abundant 
than  low-energy electrons \cite{scya+93,pani+94}.

The energy absorption is maximized by an increased momentum
during one half cycle of the pulse.
In a full laser cycle a free electron absorbs as much energy as
it  looses.
If, however, the drift momentum is increased by $p$ for just one
half of the cycle there is a net energy absorption.
Using \eq{momentum} it is given by
\begin{equation}
  \label{eq:enab}
  \Delta E=\int f(t)P(t)\,\mathrm{d}t = 
  F\,p\int\limits_{-\pi/2\omega}^{+\pi/2\omega}
  \cos(\omega t)\,\mathrm{d} t =2A\,p,
\end{equation}%
in accordance with the result in Eq.\,(\ref{eq:vmax1}).  Whereas the
absolute change of the momentum in the field, being $2A$, is
independent of the drift momentum $p$ the change of the kinetic energy
\eq{eq:enab} is proportional to it.
Rewritten in terms of energies, we obtain for the maximum energy gain 
through rescattering from 
extended systems the central result of this paper:
\begin{equation}
  \label{eq:deee}
\Delta E=4\sqrt{E_\mathrm{pond}}\sqrt{V}\,.
\end{equation}
The acceleration of the electrons depends on \emph{both}, the
laser field strength $A$ and the depth $V$ of the scattering
potential.
Interestingly, this situation is akin 
to the so-called \emph{powered swing-by\/} (or gravity-assisted
maneuver) of spacecrafts \cite{pa96}. 
There, thrust for accelerating, decelerating or redirecting  
the spacecraft is applied only for short intervals of time and
not in an oscillatory fashion like in the case of  a laser. 
However, similar to the situation considered here, 
thrust is most effectively applied when the space craft has
high momentum, which is the case at the perihelion of a 
swing-by at a planet. 

As mentioned before, for a given potential the phases
$\varphi_\i$ and $\varphi_\f$ are linked through Eq.\,(\ref{eq:p1p2}).
For the case of deep potentials, when $p\,{\gg}\,p_\i$ and
$p\,{\gg}\,A$, this equation simplifies to 
$\width\omega\approx p(\varphi_\f-\varphi_\i)$.
This fixes the difference of the phases and one may write
the momentum explicitely as
$p_\f(\varphi)=2\left[A\,p\,\cos\left(\varphi/2\right)
  \sin\left(\width\omega/2p\right)\right]^{1/2}$
with $\varphi:=\varphi_\f+\varphi_\i$ the only parameter left for optimizing
$p_\f$. For $\varphi=0$, i.\,e.\ when $\varphi_\i$ and
$\varphi_\f$ are symmetric with respect to the field maximum,
the momentum reads
\begin{equation}
  \label{eq:vfdp}
  p_\f
  =p_{\f,\mathrm{max}}
    \left[\sin\left(\frac{\pi}{2}\frac{\width}{\width_\mathrm{max}}\right)\right]^{1/2}
\end{equation}
with the maximal momentum from Eq.\,(\ref{eq:vmax1}) and the
optimal system width $\width_\mathrm{max}$ from
Eq.\,(\ref{eq:rmax1}).

\begin{figure}[t]
  \centering
\includegraphics[scale=0.5]{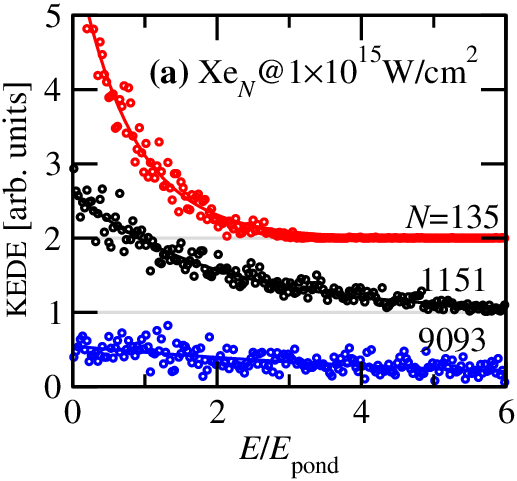}\hfill{}
\includegraphics[scale=0.5]{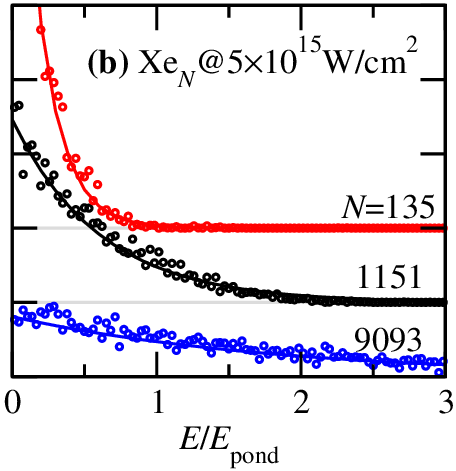}
\caption{(Color online) Kinetic energy distribution of electrons
  for two  laser pulses of 400\,fs duration and
  peak intensities of 1$\times$10$^{15}$\,W/cm$^2$ (a) and
  5$\times$10$^{15}$\,W/cm$^2$ (b), respectively.
  Shown are the results for three different cluster sizes
  Xe$_{135}$, Xe$_{1151}$ and Xe$_{9093}$, respectively.
  The first two are shifted upwards for better visibility.
  Each set of points is fitted (solid lines) by an exponential
  curve $\exp(-E/E_\mathrm{kin})$.
  Note that the energy scale is given by the corresponding
  ponderomotive energy $E_\mathrm{pond}$ which is proportional to
  the laser intensity $I$.}
\label{fig:rescatt}
\end{figure}%

To illustrate the relevance of the rescattering mechanism in extended
systems we present in the following results of microscopic
calculations for rare-gas cluster exposed to intense laser radiation
\cite{sasi+06}.  The theoretical approach describes the laser-driven
electronic nano-plasma and the ionic explosion dynamics by means of
classical equations of motion.  It has been successfully
applied to study, e.\,g., absorption mechanisms for a wide range of
clusters sizes and laser pulses by various groups
\cite{rosc+97,lajo00,isbl00,saro03}.  Figure~\ref{fig:rescatt} shows kinetic
energy distribution of electrons as obtained for xenon clusters of
different sizes and various laser pulse parameters.  The calculations
are rather expensive since the the electrons have to be propagated for
a long time (typically a few picoseconds) in order to obtain
converged results for their final kinetic energy, which otherwise
would be spoiled by the large, long-range and time-dependent
attractive potential of the exploding cluster.  All distributions show
an exponential behavior in accordance with observations from
experiments for somewhat larger xenon clusters \cite{spas+03,kukr+03}
as well as silver cluster of similar size \cite{fedo+07}.  Fitting an
exponential shape to these distributions yields parameters
$E_\mathrm{kin}$ which are listed in \tab{tab:para}.  
They strongly depend on the cluster size and the laser pulse,
but reveal a clear trend: 
The larger the cluster the faster the emitted electrons.  
This trend originates in the deeper potential for larger
clusters, i.\,e.\ larger $V$ in \eq{eq:deee}, which leads to a
stronger acceleration at rescattering. 

Note, that the energy can exceed the ponderomotive energy considerably
as seen in particular for the largest cluster Xe$_{9093}$.  Knowing
the cluster potential from the simulation we can estimate the electron
energies with the rescattering model of Eq.\,(\ref{eq:deee}) and list
them for comparison in \tab{tab:para}.  They agree surprisingly
well considering the simplicity of the model and the fact that the
electron spectrum of the microscopic calculations (\fig{fig:rescatt})
contains also all  electrons released directly.  We attribute the 
agreement to the fact that the exponential tail is due to the fast
electrons which are dominantly emitted by rescattering.  Additionally,
most of the electrons are ejected at the resonance of the cluster
\cite{kukr+03,fedo+07} where the acceleration is optimal. 
This is certainly not the case for the 
smallest cluster, Xe$_{135}$, considered here. It is almost
completely disintegrated at the time when the laser pulse reaches its
peak. What we assume to determine the parameters for rescattering 
is therefore not valid  and consequently poor is the quantitative 
prediction of the electron spectrum by rescattering for this
cluster.
\begin{table}[t]
  \caption{Electron energies in units of the
    ponderomotive energy $E_\mathrm{pond}$ from xenon clusters
    of three sizes and two different laser pulses.
    The exponential fit parameter $E_\mathrm{kin}$ from the
    microscopic calculations shown in \fig{fig:rescatt} are
    compared to the value $\Delta E$ of the rescattering model
    in \eq{eq:deee}.} 
  \centering
  \begin{ruledtabular}
    \begin{tabular}{cccccccccc}
      cluster size $N$ 
      && \multicolumn{2}{c}{135} && \multicolumn{2}{c}{1151} && \multicolumn{2}{c}{9093} \\
      laser pulse && (a) & (b) && (a) & (b) && (a) & (b) \\
      \begin{tabular}[b]{r}microscopic result \\ 
        $E_\mathrm{kin}$/$E_\mathrm{pond}$\end{tabular}
      && 0.78 & 0.27 && 1.98 & 0.60 && 5.43 & 1.74 \\
      \begin{tabular}[b]{r}rescattering model \\ 
        $\Delta E$/$E_\mathrm{pond}$\end{tabular}
      && 0.19 & 0.09 && 1.53 & 0.65 && 3.68 & 1.58 \\
    \end{tabular}
  \end{ruledtabular}
  \label{tab:para}
\end{table}%

\begin{figure}[t]
  \centering
  \includegraphics[width=0.3\textwidth]{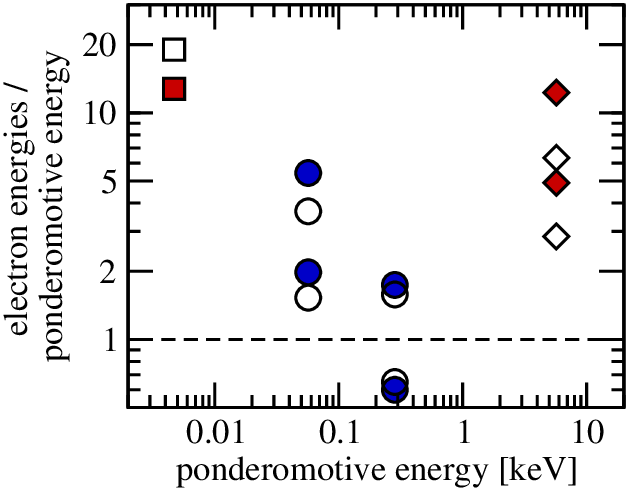}
  \caption{Electron energies as a function of the ponderomotive
    energy $E_\mathrm{pond}$ from various clusters.
    The filled symbols show $E_\mathrm{kin}$ for experiments
    (red) and microscopic calculations (blue):
    Ag$_{1000}$ (square, \cite{fedo+07}), 
    Ar$_{1700}$ and Ar$_{33000}$ (diamonds, \cite{chpa+02}),
    Xe$_{1151}$ and Xe$_{9093}$ (circles, \fig{fig:rescatt}).
    The corresponding estimates $\Delta E$ from the rescattering
    model (\ref{eq:deee}) are shown by open symbols.
  }
  \label{fig:exp}
\end{figure}

Similar to the calculated electron energy distributions
experimental spectra can be characterized by constants
$E_\mathrm{kin}$ quantifying the exponential decay.  
They are shown for measurements
\cite{fedo+07,chpa+02} of various cluster 
along with the theoretical results discussed above
in \fig{fig:exp}.
Clearly, they are larger than the ponderomotive energy, i.\,e., above
the dashed line in \fig{fig:exp}.  Note that corresponding data for
atoms \cite{pani+94} are below this line.  Even in cases where one
observes a plateau energetic electrons from atoms are much less
likely than for extended systems such as clusters.

In contrast to the microscopic calculation we generally neither know
the charge nor the radius of the cluster at the peak of the laser
pulse.
However, there are two exceptions: (i) the pulse is very short
\cite{chpa+02} or (ii) the delay of a dual pulse
is adjusted to induce resonant ionization \cite{fedo+07}.
In both cases one can roughly estimate the unknown cluster
parameters  which determine width $\width$ and depth $V$ of the 
extended  potential for rescattering.
In case (i) one can neglect the cluster expansion, the radius of the
scattering potential is the initial cluster radius $R$.  The charge
$Q$ can be estimated from a simple over-the-barrier model $Q=F R^2$
\cite{sa06}.  Hence, with the potential depth $V{=}\,3Q/2R$ of a
homogeneously charged sphere we can by means of \eq{eq:deee} determine
the electron energies (open diamonds in \fig{fig:exp}) in reasonable
agreement with the experimental values.  In particular the ratio between
the larger and the smaller cluster size is well reproduced.  Note that
the experimental signal is a sum over clusters of different size and
the laser focus.  This may be crucial \cite{issa+06} and agreement
with a simple model on the absolute scale cannot be expected.
For case (ii) we assume that the cluster expands homogeneously
over the delay between the (short) double pulses.
Thus one has at resonance $Q/R^3\equiv\omega^2$ \cite{saro03}.  
The charge $Q$ can be roughly assessed from
measured final ion charges \cite{dofe+05}.  
Assuming an average ion charge of two we estimate a value (open
square in \fig{fig:exp}) slightly above the experimental one.

The experimental and numerical examples of rare-gas clusters
demonstrate that the simple rescattering mechanism for extended
potentials we have introduced can provide considerable insight
into complicated many-body dynamics as it occurs in these
clusters including semi-quantitative predictions.
However, the analytical results from Eqs.\,(\ref{eq:vdve}) and
(\ref{eq:p1p2}) are far more general than the example of rare
gas clusters may suggest and should, e.\,g., also describe
non-linear absorption of laser energy in quantum dots. 
Moreover, rescattering in extended systems does not only provide
conditions for optimum energy absorption from the light, it also 
answers the opposite question: 
Given a certain rescattering potential  and  laser pulse, what
is the maximum velocity of a particle which can be brought to
rest (remains sticking in the scattering system) under the  
combined action of laser and potential?  
This situation is also described by \eq{eq:vmax} if one
interchanges the indices ``f'' and ``i'' and sets $p_\f=0$. 
The result is given by \eq{eq:vmax1}, but now
$p_{\i,\mathrm{max}} \approx 2\sqrt{Ap}$ stands for the maximum
initial momentum an electron can have and still will stick to
the extended potential being most efficiently decelerated by the
potential and the light.  
This constitutes  non-linear photo-association in extended systems.

We acknowledge support from the Kavli Institute for Theoretical
Physics (KITP) in Santa Barbara, where this work has been started.

\end{document}